# An Algebraic Formalization of the GoF Design Patterns

Paolo Bottoni[1], Esther Guerra[2], Juan de Lara[3]
[1] Computer Science Department, "*Sapienza*" Università di Roma (Italy)
[2] Computer Science Department, Universidad Carlos III de Madrid (Spain)
[3] Computer Science Department, Universidad Autónoma de Madrid (Spain)

**Abstract**
This document reports on the use of an algebraic, visual, formal approach to the specification of patterns for the formalization of the GoF design patterns. The approach is based on graphs, morphisms and operations from category theory and exploits triple graphs to annotate model elements with pattern roles. Being based on category theory, the approach can be applied to formalize patterns in different domains. Novel in our proposal is the possibility of describing (nested) variable submodels, inter-pattern synchronization across several diagrams (e.g. class and sequence diagrams for UML design patterns), pattern composition, and conflict analysis.

## *1. Introduction*

Patterns [Alex77] are increasingly used in the definition of software frameworks [GoF94], as well as in Model Driven Development, to indicate parts of required architectures [AN04], drive code refactoring [K04], or build model-to-model transformations [BJP04]. The full realisation of their power is however hindered by the lack of a standard formalization of the notion of pattern. Presentations of patterns are typically given through natural language, to explain their motivation, context and consequences; programming code, to show usages of the pattern; and diagrams, to communicate their structure and behaviour.

Several researchers have indicated the limitations of the current semi-formal devices for pattern definition – generally based on domain modelling languages, such as UML for design patterns, or Coloured Petri Nets for workflow – and research is active to propose rigorous formalisms, methodologies and languages for pattern definition in specific domains, as well as to propose general models of patterns.

The availability of formalisms will make common practices involving patterns, such as pattern discovery, pattern enforcement, pattern-based refactoring, etc., simpler and amenable to automation, and open new perspectives for pattern composition and analysis of pattern consequences.

In this document, we report on the use of an algebraic approach to the specification of patterns [BGL09, BGL10] for the formalization of the GoF patterns. The approach is based on graphs, morphisms and operations from category theory and exploits triple graphs to annotate model elements with pattern roles. Being based on category theory, the approach can be applied to formalize patterns in different domains. Novel in our proposal is the possibility of describing (nested) variable submodels, inter-pattern synchronization across several diagrams (e.g. class and sequence diagrams for UML design patterns), pattern composition, and conflict analysis.

The rest of the document is organized as follows. Section 2 provides the fundamentals of our formalization, Section 3 presents the formalization of the GoF patterns, and Section 4 draws some conclusions.

## 2. Formalization of Patterns

Here we just give an introduction to the basics of our formalization, the reader is urged to consult [BGL09, BGL10] for full details.

In its simplest form, a pattern consists of one structure, called *root*, with the mandatory part that any pattern realization must contain, and a number of *variable* parts or variability regions defining additional structures that can be replicated several times for each instance of the root [BGL09]. In order to represent root and variable parts, we use the symbolic graphs proposed in [O08], where the set of data nodes is replaced by a finite set of sorted variables, and a formula γ constrains the possible values taken by the variables. For simplicity, we first present them without role annotations.

Variable parts can be nested: a nested part can only be instantiated by adding structures to an instance of its parent. In addition, for each variable part, a variable name with values ranging on *integer* is used in equations restricting the allowed number of its replicas. Equations may contain relations between the allowed replicas of different variable parts, and the number of times a pattern can be instantiated in a model can be restricted by equations on the variability of the root. If the set of equations has no solution in the natural numbers, then the pattern cannot be instantiated.

**Def. 1. Pattern.** A pattern is a construct *VP=(P, root, Emb, name, var)*, where:
- $P=\{V_1, ..., V_n\}$ is a finite set of non-empty graphs, where each $V_i$ is called *variable part*,
- *root* ∈ P is a distinguished element of P, also called the fixed part,
- *Emb* is a set of morphisms $v_{i,j}: V_i \rightarrow V_j$ with $V_i, V_j \in P$, s.t. it spans a tree rooted in *root* with all graphs $V_i \in P$ as nodes and the morphisms $v_{i,j} \in Emb$ as edges,
- name: P→L is an injective function assigning each variable part a *name* from a set of variables *L*, of sort N,
- var ⊆ $T_{AlgIEq}$(name(P)) is a set of equations governing the number of possible instantiations of the variable parts, using variables in name(P) ⊆ L, arithmetic operations, and the <, ≤, =, >, ≥ relation symbols. We call this signature "*Algebraic Inequalities*" ($\Sigma_{AlgIEq}$); $T_{AlgIEq}$(name(P)) is the term-algebra with variables in name(P).

In figures, we represent variability regions enclosed in polygons. Two nested variable parts $v_{i,j}: V_i \rightarrow V_j$ are depicted through containment.

The semantics of a pattern is given by the set of all valid expansions of its variability regions.

**Def. 2. Expansion**. The expansion set *EXP(VP)* of a pattern *VP* is given by the set of colimits $\{C_i\}$ of all possible diagrams α obtained by replicating the graphs in *P*, and the morphisms in *Emb*, s.t.: (i) the diagram α is consistent with the morphisms in *Emb*, which means that if $V_i \rightarrow V_j$ is included in α, then there is a morphism $v_{i,j}:V_i \rightarrow V_j \in Emb$; and (ii) the number of replicas in each path from *root* to $C_i$ satisfies the equations in *var*.

A model satisfies a variable pattern when some pattern expansion is found in the model.

**Def. 3. Satisfaction and Semantics**. Given a pattern *VP* and a graph *G*, *G* satisfies *VP*, written G ⊨ VP, iff ∃$C_i$→G injective with $C_i$∈EXP(VP). The semantics of VP, SEM(VP)={G | G ⊨ VP} is given by the set of all graphs that satisfy it.

We annotate nodes in the graph with the roles they play in the pattern, thus using triple graphs instead of graphs as objects in the set P. A triple graph [S94] consists of two separate graphs (called source and target) related through a correspondence graph. In our approach, the source represents a model in a given domain-specific language (e.g. UML), while the target contains nodes with the different roles the elements can take. The assignment of roles to elements is made through the nodes in the correspondence graph, which have morphisms to source and target nodes.

Patterns may include contextual conditions needed for their correct application, expressed as graph application conditions [EEH06]. To a certain point, they help in expressing the intent and consequences of the pattern. A *pattern with invariants* is a pattern with sets PC($V_i$) of pattern constraints defined over any of its variable parts $V_i$. An atomic constraint over $V_i$, is given by a constraint $V_i \rightarrow X \rightarrow C_j$, and consists of one *premise* graph X (related to the variable part $V_i$ it constrains) and a set of *consequences* $C(X)=\{X \rightarrow C_j\}_{j\in J}$. As in logic, the intuition behind constraint satisfaction is: if the premise graph X is found in a model, then some of the consequence graphs $C_j$ have to be found as well. Note that if the index set *J* is empty, then we have a negative constraint (NAC), forbidding X to be found. This kind of constraint is represented visually as a crossed out polygon.

Patterns are usually given by the synchronization of two or more diagrams. For this purpose, we allow one primary pattern (usually expressing the structure, e.g., a class diagram in UML), synchronized with one or more secondary patterns (usually expressing behaviour, e.g., a collaboration diagram in UML). The synchronization is performed by identifying common elements in the variable parts of each pattern, and is formalized by a synchronization graph [BGL09].

## *3. Formalization of the GoF Design Patterns*

This section presents the formalization of all GoF patterns. Most patterns are characterized by a structural part (class diagram) and a behavioural part (sequence diagram). For reference, we just provide the intent of each pattern (taken literally from [GoF94]), the reader is directed to [GoF94] for further details.

Some patterns contain also invariants, which are contextual requirements needed for a correct application of the pattern. We make some use of "notes", but the behavioural specification of some methods is given by activity diagrams. For clarity, all patterns are presented in concrete syntax. For illustration purposes, the Proxy is also presented in abstract syntax. All textual elements (names of classes, operations, etc.) starting by upper case are variables, unless otherwise stated.

### **3.1. Abstract Factory**
<u>*Intent:*</u> Provide an interface for creating families of related or dependent objects without specifying their concrete classes.

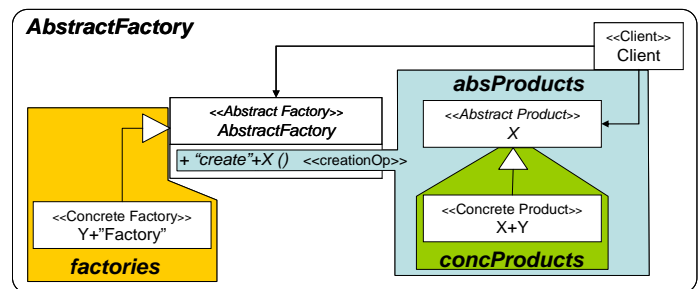

Fig. 1: Structure of the Abstract Factory

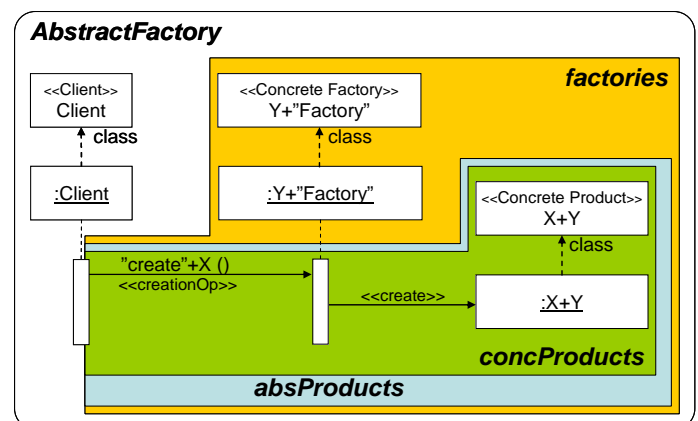

Fig. 2: Collaboration for the Abstract Factory

## 3.2. Adapter
*Intent:* Convert the interface of a class into another interface clients expect. Adapter lets classes work together that couldn't otherwise because of incompatible interfaces.

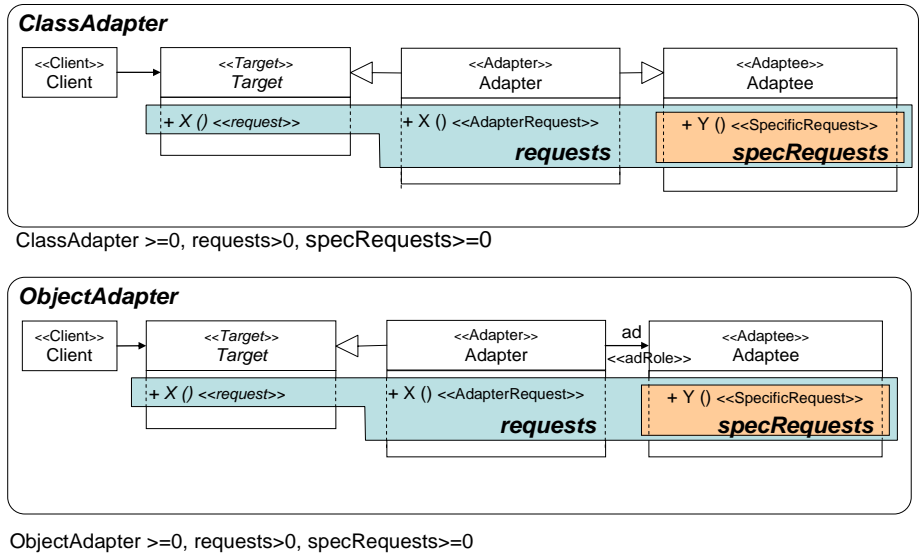

Fig. 3: Structure of Class and Object Adapter.

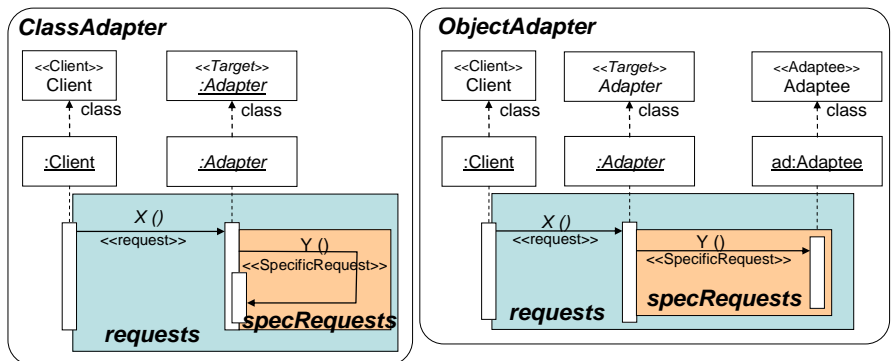

Fig. 4: Collaborations for Class and Object Adapter.

## 3.3. Bridge
*Intent:* Decouple an abstraction from its implementation so that the two can vary independently.

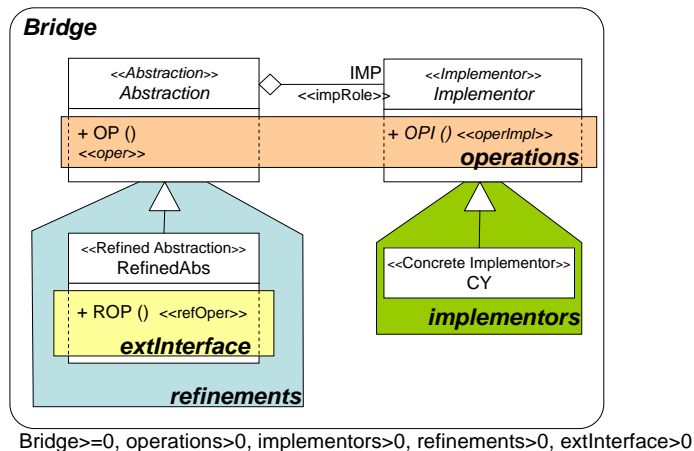

Fig. 5: Structure of Bridge.

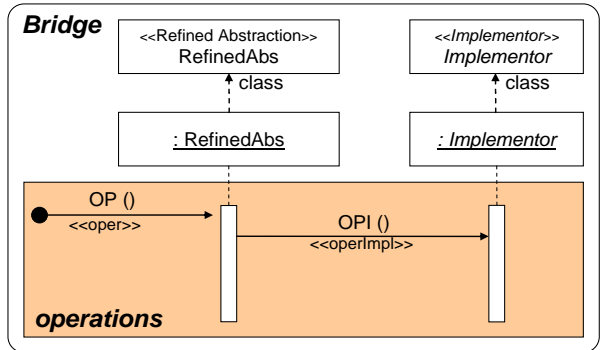

Fig. 6: Collaboration for Bridge.

## 3.4. Builder

***Intent:*** Separate the construction of a complex object from its representation so that the same construction process can create different representations.

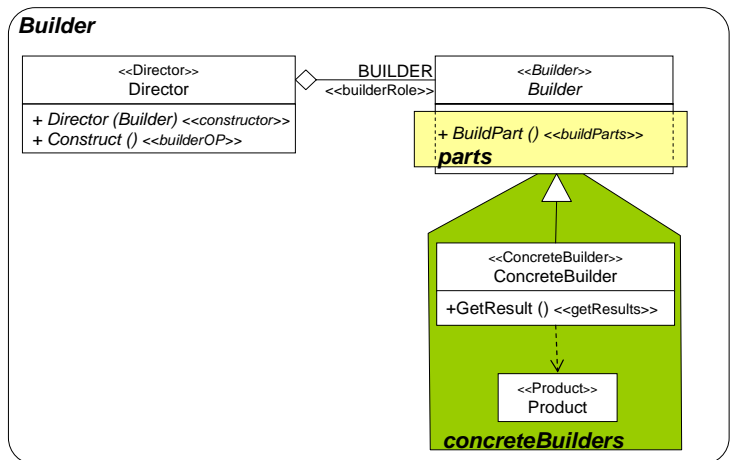

Builder>=0, concreteBuilders>0, parts>0

Fig. 7: Structure of Builder.

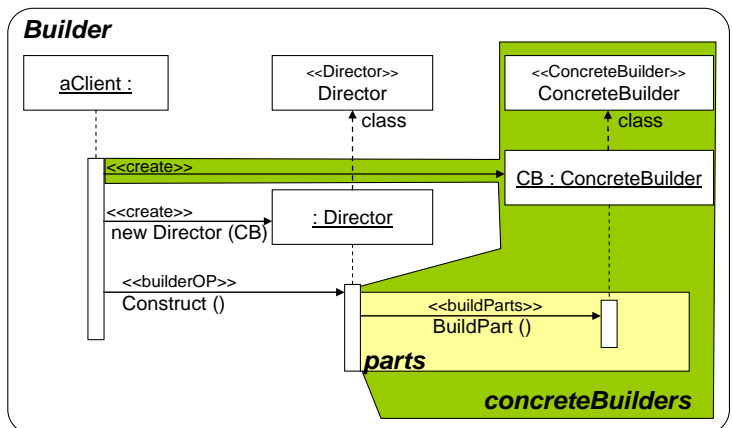

Fig. 8: Collaboration for Builder.

## 3.5. Chain of Responsibility

***Intent:*** Avoid coupling the sender of a request to its receiver by giving more than one object a chance to handle the request. Chain the receiving objects and pass the request along the chain until an object handles it.

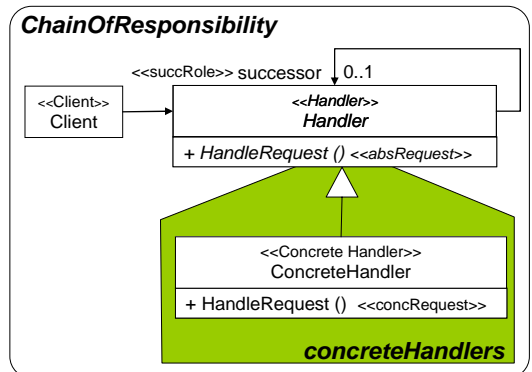

Fig. 9: Structure of Chain of Responsibility.

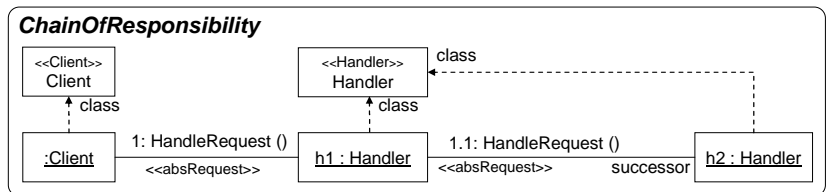

Fig. 10: Collaboration for Chain of Responsibility.

## 3.6. Command

***Intent:*** Encapsulate a request as an object, thereby letting you parameterize clients with different requests, queue or log requests, and support undoable operations.

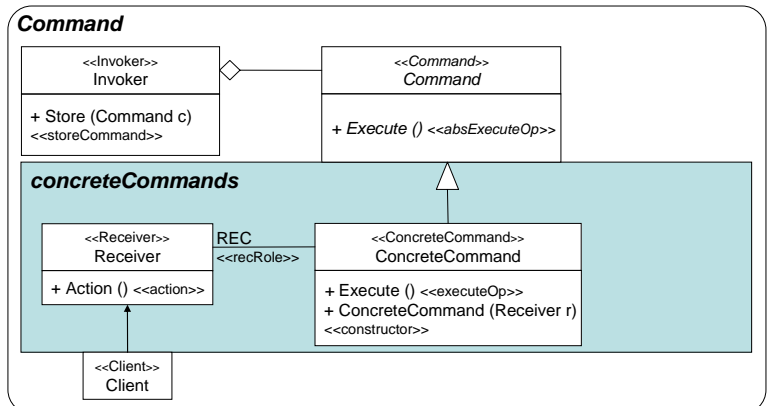

Fig. 11: Structure of Command.

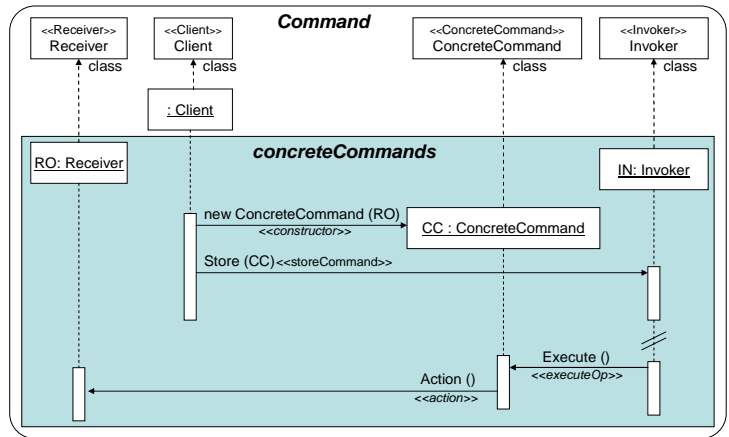

Fig. 12: Collaboration for Command.

## 3.7. Composite

***Intent:*** Compose objects into tree structures to represent part-whole hierarchies. Composite lets clients treat individual objects and compositions of objects uniformly.

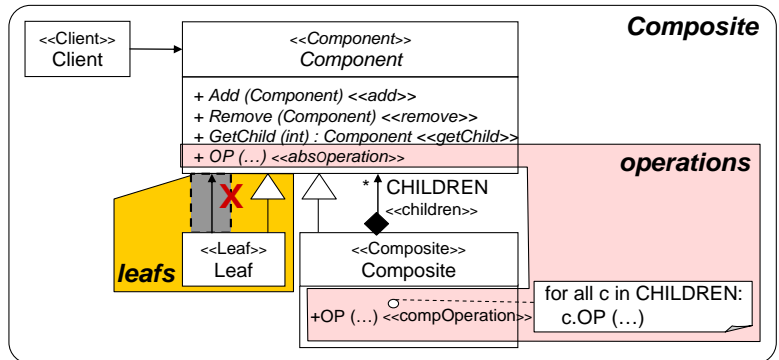

Composite>=0, operations>0, leafs>0

Fig. 13: Structure of Composite.

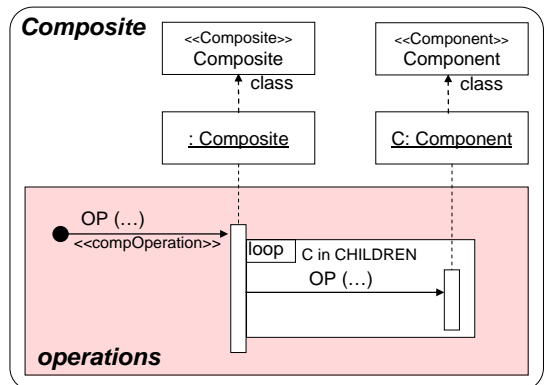

Fig. 14: Collaboration for Composite.

## 3.8. Decorator

***Intent:*** Attach additional responsibilities to an object dynamically. Decorators provide a flexible alternative to subclassing for extending functionality.

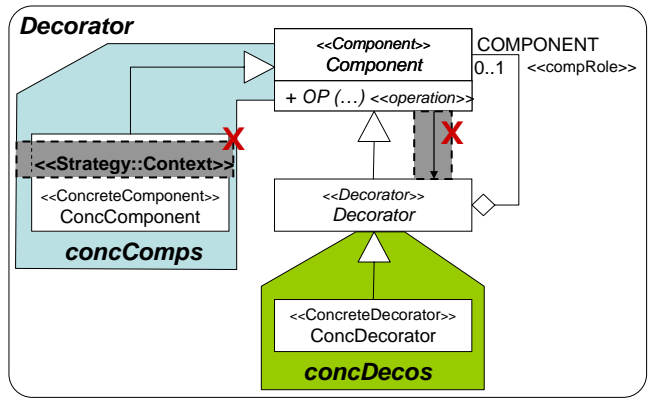

Decorator>=0, concComps>0, concDecos>0
Fig. 15: Structure of Decorator.

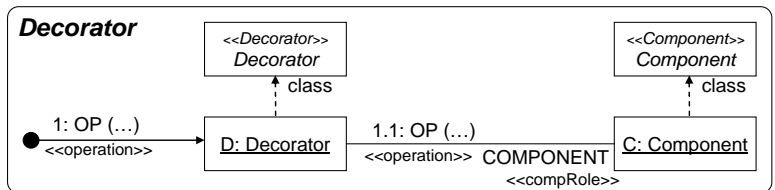

Fig. 16: Collaboration for Decorator.

## 3.9. Facade

*Intent:* Provide a unified interface to a set of interfaces in a subsystem. Facade defines a higher-level interface that makes the subsystem easier to use.

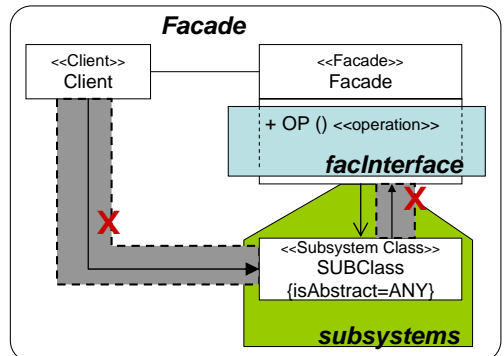

Facade>=0, facInterface>0, subsystems>0
Fig. 17: Structure of Facade.

## 3.10. Factory Method

*Intent:* Define an interface for creating an object, but let subclasses decide which class to instantiate. Factory Method lets a class defer instantiation to subclasses.

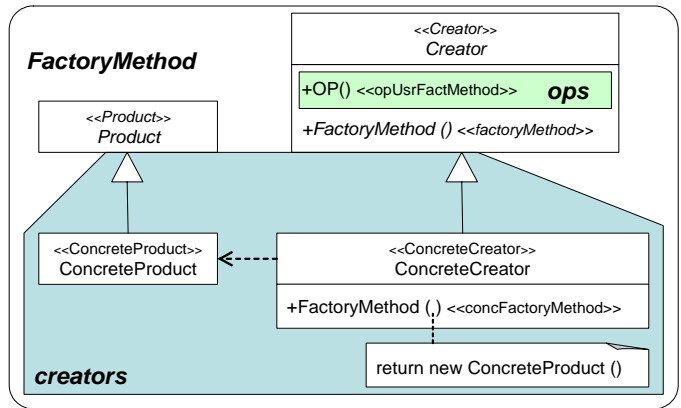
Fig. 18: Structure of Factory Method.

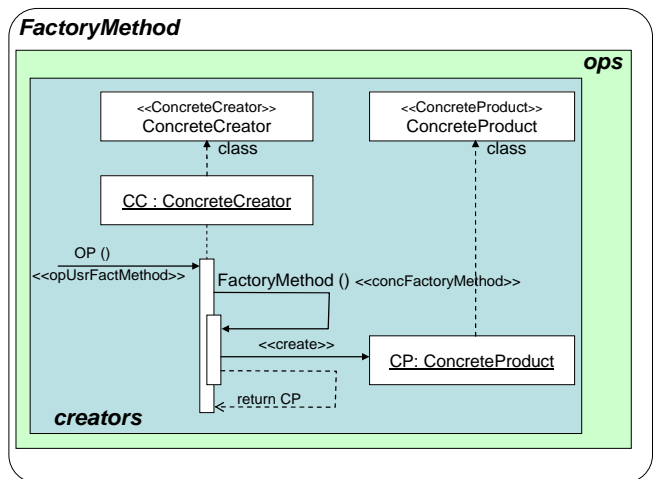
Fig. 19: Collaboration for Factory Method.

## 3.11. Flyweight
*Intent:* Use sharing to support large numbers of fine-grained objects efficiently.

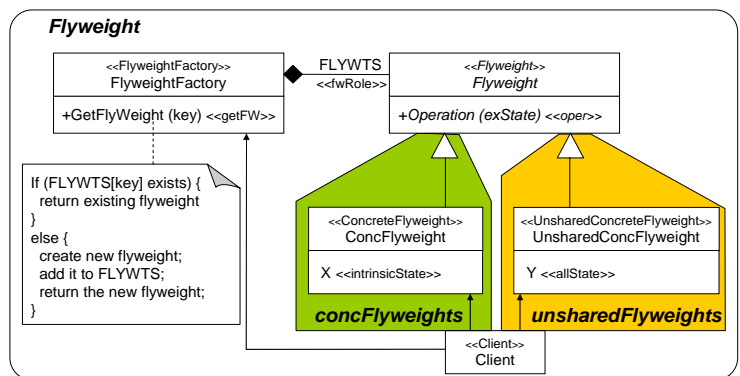
Fig. 20: Structure of Flyweight.

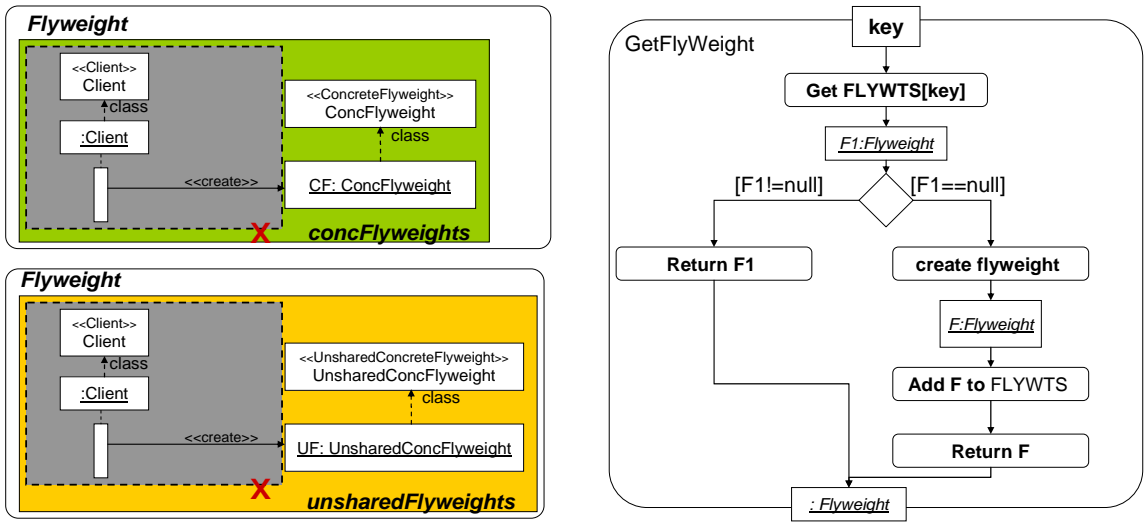

Fig. 21: Collaborations for Flyweight (left). Activity Diagram (right)

## 3.12. Interpreter

***Intent:*** Given a language, define a representation for its grammar along with an interpreter that uses the representation to interpret sentences in the language.

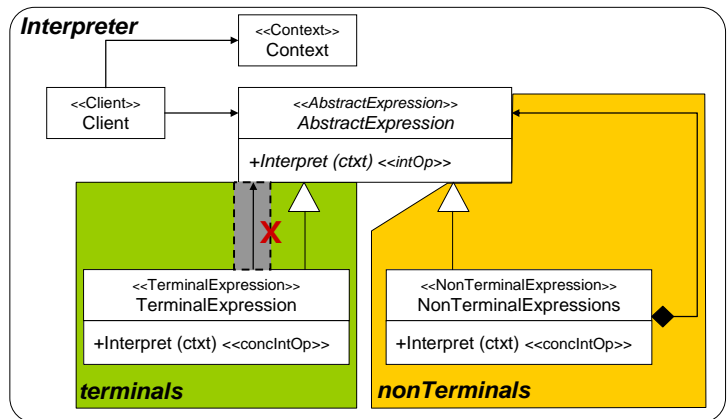

Interpreter>=0, terminals>0, nonTerminals>0

Fig. 22: Structure of Interpreter.

## 3.13. Iterator

***Intent:*** Provide a way to access the elements of an aggregate object sequentially without exposing its underlying representation.

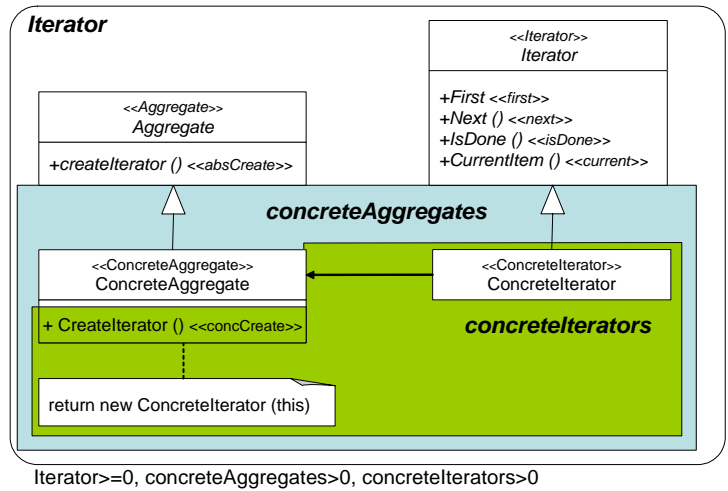
Iterator>=0, concreteAggregates>0, concreteIterators>0
Fig. 23: Structure of Iterator.

## 3.14. Mediator

***Intent:*** Define an object that encapsulates how a set of objects interact. Mediator promotes loose coupling by keeping objects from referring to each other explicitly, and it lets you vary their interaction independently.

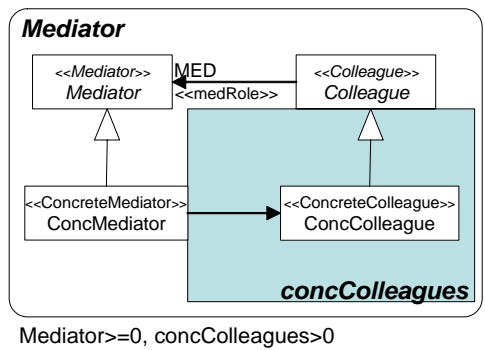
Mediator>=0, concColleagues>0
Fig. 24: Structure of Mediator.

## 3.15. Memento

***Intent:*** Without violating encapsulation, capture and externalize an object's internal state so that the object can be restored to this state later.

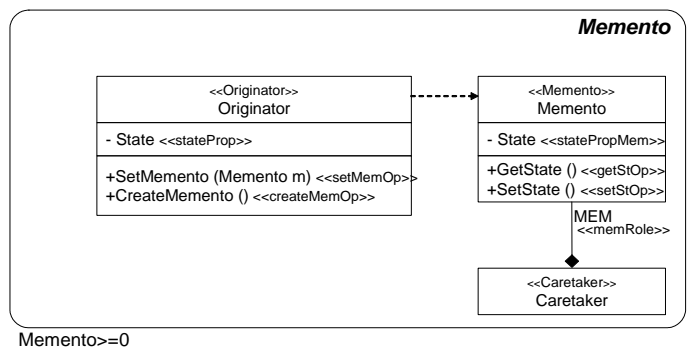
Memento>=0
Fig. 25: Structure of Memento.

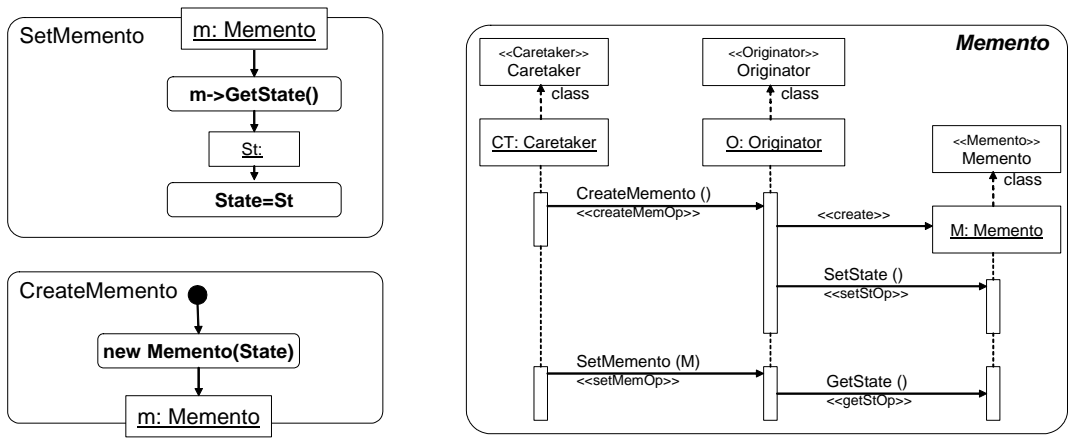

Fig. 26: Activity Diagrams for Memento (left). Collaboration for Memento (right).

## 3.16. Observer

***Intent:*** Define a one-to-many dependency between objects so that when one object changes state, all its dependents are notified and updated automatically.

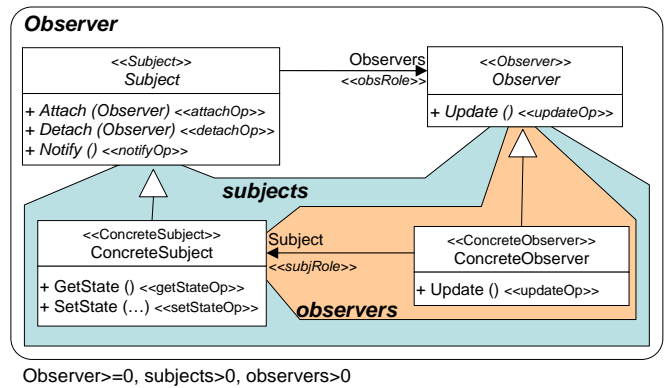

Observer>=0, subjects>0, observers>0

Fig. 27: Structure of Observer.

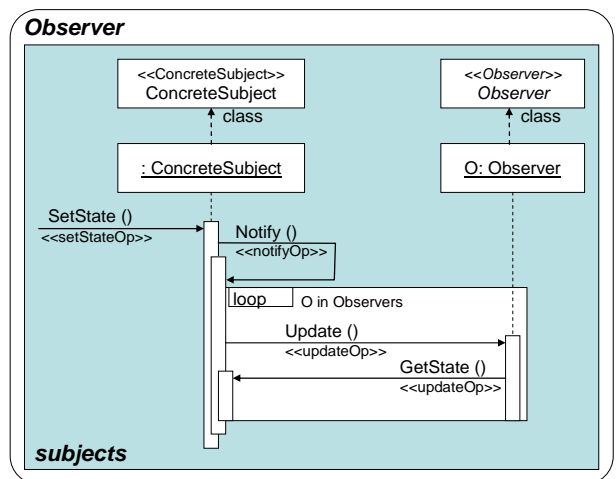

Fig. 28: Collaboration for Observer.

## 3.17. Prototype

***Intent:*** Specify the kinds of objects to create using a prototypical instance, and create new objects by copying this prototype.

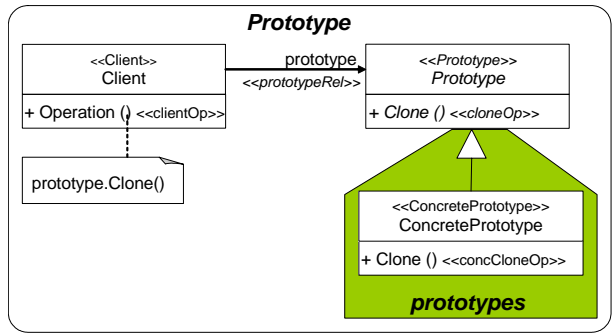
Prototype>=0, prototypes>0
Fig. 29: Structure of Prototype.

## 3.18. Proxy

*Intent:* Provide a surrogate or placeholder for another object to control access to it.

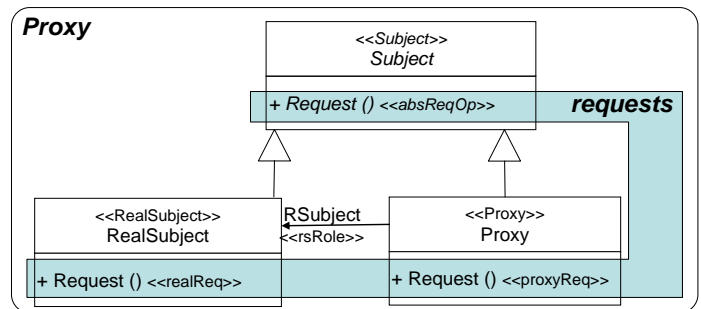
Proxy>=0, requests>0
Fig. 30: Structure of Proxy.

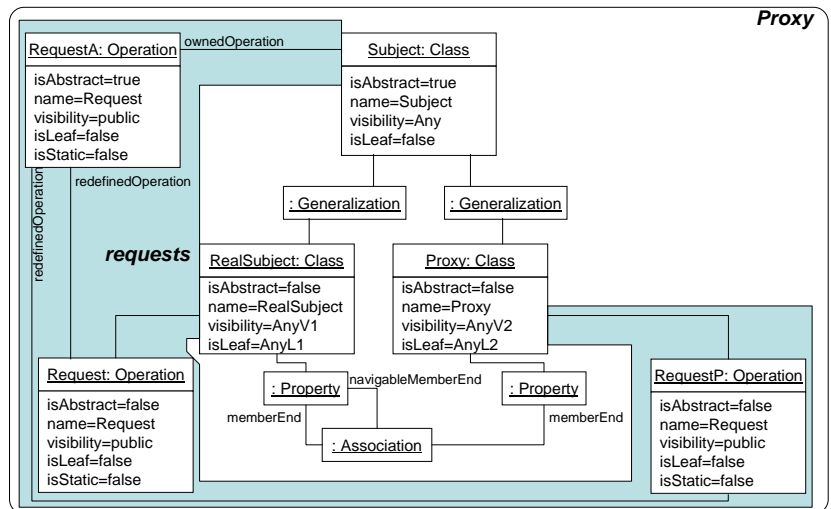
Fig. 31: Structure for Proxy (Abstract Syntax).

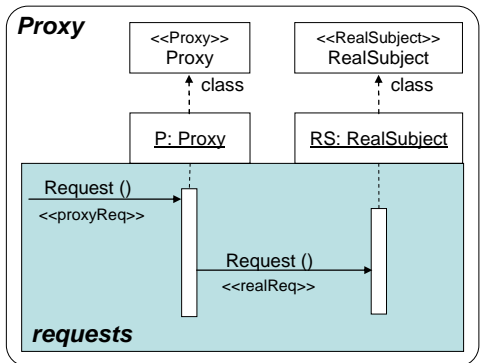
Fig. 32: Collaboration for Proxy.

## 3.19. Singleton
*Intent:* Ensure a class only has one instance, and provide a global point of access to it.

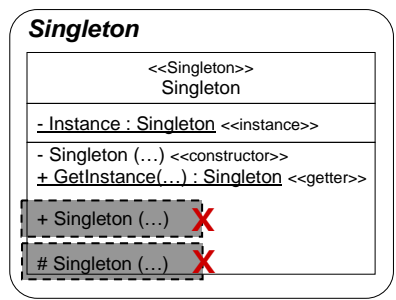
Fig. 33: Structure of Singleton.

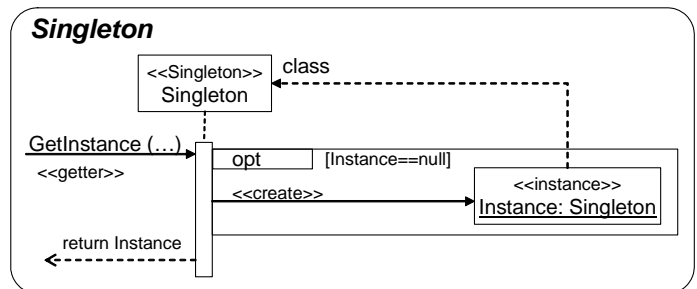
Fig. 34: Collaboration for Singleton.

## 3.20. State
*Intent:* Allow an object to alter its behaviour when its internal state changes. The object will appear to change its class.

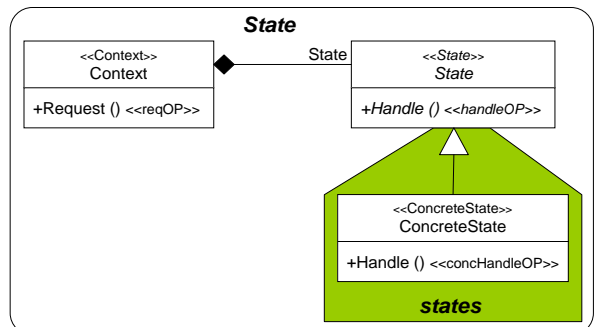
Fig. 35: Structure of State.

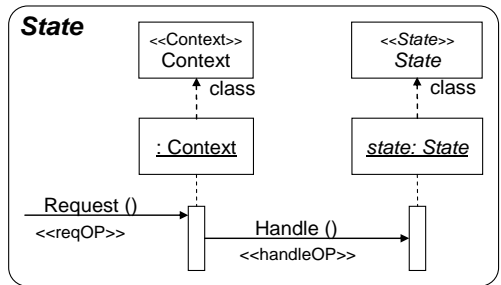

Fig. 36: Collaboration for State.

## 3.21. Strategy

*Intent:* Define a family of algorithms, encapsulate each one, and make them interchangeable. Strategy lets the algorithm vary independently from clients that use it.

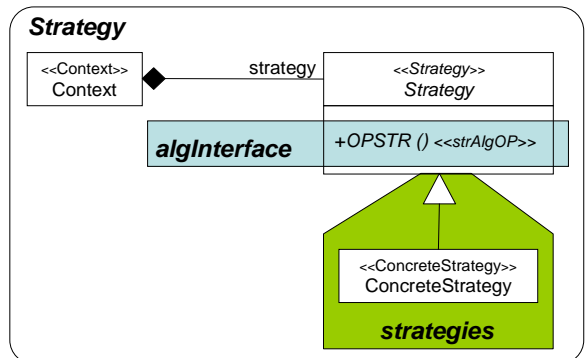

Strategy>=0, strategies>0, algInterface>0

Fig. 37: Structure of Strategy

## 3.22. Template Method

*Intent:* Define the skeleton of an algorithm in an operation, deferring some steps to subclasses. Template Method lets subclasses redefine certain steps of an algorithm without changing the algorithm's structure.

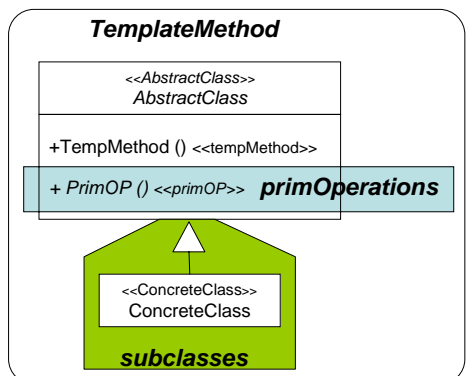

TemplateMethod>=0, primOperations>0, subclasses>0

Fig. 38: Structure of Template Method

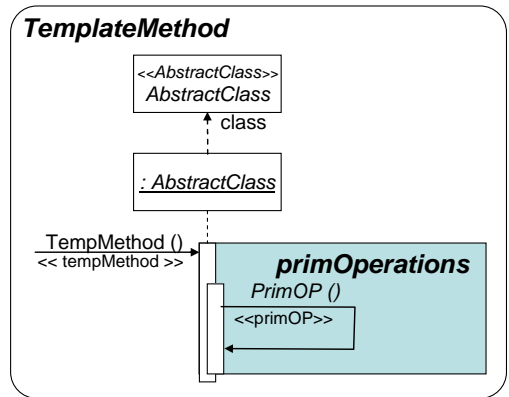

Fig. 39. Collaboration for Template Method

## 3.23. Visitor

*Intent:* Represent an operation to be performed on the elements of an object structure. Visitor lets you define a new operation without changing the classes of the elements on which it operates.

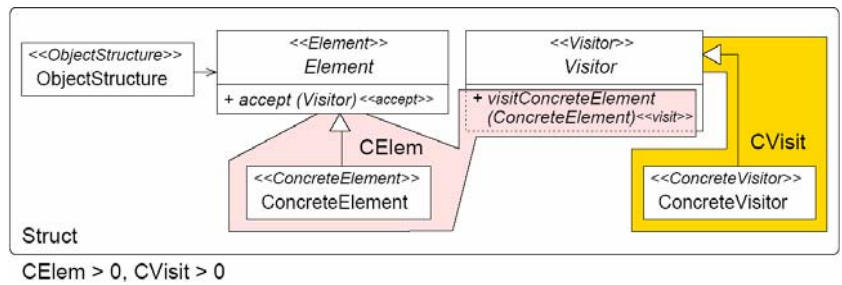

Fig. 40. Structure of Visitor.

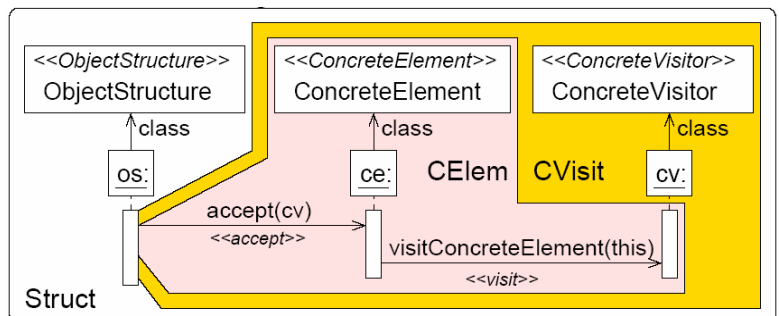

Fig. 41. Collaboration for Visitor.

## *4. Conclusions*

In this paper, we have presented an algebraic formalization of all GoF design patterns. The formalization allows specifying the synchronization of an structuring pattern (e.g. a class diagram) and several interaction patterns (e.g. collaborations). Our formalization also allows attaching invariants (similar to graph application conditions) to the variable parts of the patterns, hence expressing the intent.

Some extensions that may add more flexibility to our formalization would be the following ones. First, we could allow relating two independent variability regions through edges. This would mean replicating such edge for each replica of both regions. This would demand extending *Emb* to be a graph instead of a tree, but would complicate the definition of the expansion set. Second, the replication of variability regions is "in width", but it could also be interesting to explore replications "in length", e.g., to construct paths of arbitrary length. Finally, we could define more sophisticated logics to control the instantiations of the variability regions, and provide application conditions between elements of different variability regions.